\documentclass[12pt,journal,draftcls,letterpaper,onecolumn]{IEEEtran}
\usepackage{amssymb}
\usepackage{amsmath}
\usepackage{amsfonts}
\usepackage{epsfig}
\usepackage{algorithm}
\usepackage{graphicx}
\usepackage{subfigure}
\usepackage{bm}
\usepackage{amsthm}
\usepackage{multirow}
\usepackage{tikz}
\usetikzlibrary{arrows.meta}
\usetikzlibrary{bending}
\usetikzlibrary{topaths}

\allowdisplaybreaks

\begin{document}

\title{A Hybrid Frame Structure Design of OTFS for Multi-tasks Communications}
\author{\IEEEauthorblockN{Pu~Yuan, Jin~Liu, Dajie~Jiang, Fei~Qin}
\IEEEauthorblockA{vivo Mobile Communications Co.,Ltd., Beijing, China 100015\\}
Email: \{yuanpu, jin.liu, jiangdajie, qinfei\}@vivo.com}
\maketitle

\begin{abstract}
Orthogonal time frequency space (OTFS) is a promising waveform in high mobility scenarios for it fully exploits the time-frequency diversity using a discrete Fourier transform (DFT) based two dimensional spreading. However, it trades off the processing latency for performance and may not fulfill the stringent latency requirements in some services. This fact motivates us to design a hybrid frame structure where the OTFS and Orthogonal Frequency Division Multiplexing (OFDM) are orthogonally multiplexed in the time domain, which can adapt to both diversity-preferred and latency-preferred tasks. As we identify that this orthogonality is disrupted after channel coupling, we provide practical algorithms to mitigate the inter symbol interference between (ISI) the OTFS and OFDM, and the numerical results ensure the effectiveness of the hybrid frame structure.
\end{abstract}


\section{Introduction}
Delay-Doppler domain modulation (DDM) is an emerging technique that is potentially a candidate feature for the next generation wireless communications. Unlike conventional OFDM, OTFS multiplexes the modulation symbols in the delay-Doppler domain instead of the time-frequency domain, resulting in a clear input-output relationship that can be resolved to reflect the channel coupling. One well-known variant of the DDM is the orthogonal time frequency space (OTFS) \cite{Hadani2017otfs}, which employs an ISFFT (inverse symplectic finite Fourier transform) before the OFDM modulator to implement the resource mapping in the delay-Doppler domain.

The key feature of OTFS is the two dimensional spreading via the ISFFT, which provides an opportunity for OTFS to achieve full diversity in both time and frequency. However, the cost is the relatively long latency, large hardware buffer, and the computational complexity for a frame-based detection. While being granted better spectrum and energy efficiency by removing the per-symbol cyclic prefix (CP), the OTFS suffers from inter symbol interference (ISI) which is prevented from using the single-tap equalizer. The existing detection schemes such as the message passing algorithm (MPA) \cite{Raviteja2019embedded} \cite{Raviteja2018interference} and the expectation propagation algorithm (EPA) \cite{9827946} usually exhibit high complexity or the complexity grows rapidly as the frame size increases, which may prevent practical use in certain scenarios with large frame size. A series of alternative symbol detection schemes have been proposed to alleviate the complexity, but all of them trade off for overhead. The method proposed in \cite{Raviteja2018lowcomplexity} and \cite{Qi2022iterative} sacrifices the spectrum efficiency by attaching zero-padding (ZP) in delay-Doppler domain symbols to mitigate the inter Doppler interference, then low complexity iterative RAKE decision feedback equalizer (IR-DFE) can be utilized for symbol detection. However, the ZP overhead could be reduced by implanting the pilot pulse or sequence \cite{yuan2022lowpapr} for channel estimation in the padding area. CP-OTFS is discussed in \cite{9303350} where per-OFDM-symbol CP is appended to enable symbol-by-symbol detection with low latency and complexity. In \cite{DuJoint} a precoding approach is utilized to alleviate the detection complexity at the cost of additional channel state information (CSI) signaling.

The OTFS modulation is naturally not suitable for small package transmission since the performance relies on larger frame size to enjoy better delay and Doppler resolution and time frequency diversity. Therefore, the aforementioned techniques still face the processing delay problem. 
5G NR now supports different waveforms such as CP-OFDM and DFT-S-OFDM for time division multiplexing. Users who support both waveforms can switch the waveform at either time slot or symbol level. This inspires us the following idea. If solely the OTFS waveform is not the optimal choice for certain scenarios, we might consider a frame structure to synthesis the OTFS modulation and OFDM together, allowing each of them to play to their strengths. The OTFS modulation is utilized to send large package with extra reliability and no stringent latency requirements, while the OFDM is used to send small package with less reliability and low latency.

\section{The Transmit Waveform Design}
In this section, we demonstrate the design principle of the proposed frame structure and the discrete time representation of the transmit signal. Since the delay-Doppler and the time-frequency domain are inter-convertible using ISFFT and its inverse, we can develop a simple implanting scheme based on the property of the Fourier transform to achieve an orthogonally multiplexed OTFS and OFDM symbols.

We inherit the conventional notation in 5G NR and denote the two dimensions of a time-frequency resource grid as OFDM symbols and sub-carriers, respectively. We use bold capital letters for matrices, bold lowercase letters for vectors, and italic lowercase letters for scalars. We denote $\otimes$, $\circ$ and $\oslash$ as the Kronecker product, Hadamard product and Hadamard division respectively. 

Suppose a a frequency domain sequence $\mathbf{x}$ is the DFT of a time domain sequence $\mathbf{s}$. The time domain interpolation with $i-1$ zeroes is equivalent to the frequency domain repetition $i$ times, which is illustrated in (\ref{eq:eq0}).
\begin{align}
\mathrm{IDFT}&([s_0, \underbrace{0, ...}_{i-1}, s_1, 0, ..., s_{l-1}, 0, ...]) \nonumber \\
& = [\underbrace{x_0, ..., x_{l-1}, x_0, ..., x_{l-1}, ..., x_0, ..., x_{l-1}}_{\mathrm{Duplicated\ for\ i\ times.}}]
\label{eq:eq0}
\end{align}

Therefore, by repeating the modulated symbols of OTFS in the delay or Doppler dimensions $i$ times, we let the OTFS only occupy a fraction of $\frac{1}{i}$ of the OFDM symbols or sub-carriers within a frame, while the leftovers, i.e., the unoccupied OFDM symbols or sub-carriers, can be utilized by the OFDM for resource mapping.

\subsection{Frame Structure and Transmit Signal}
We mainly consider the Doppler dimension repetition and leave idle symbols for the OFDM transmission for the following reasons. Firstly, as the OFDM user carries out the demodulation in a symbol-by-symbol manner, and several OFDM symbols may belong to the same OFDM slot, it is preferred to keep the continuity of the OFDM symbols to reduce processing latency. Secondly, in this case the inter-carrier-interference (ICI) between the OTFS and OFDM can be avoided, while the ISI can be efficiently mitigated by the proposed receiving process. 

\begin{figure}
\centering
\includegraphics[width=0.45\textwidth]{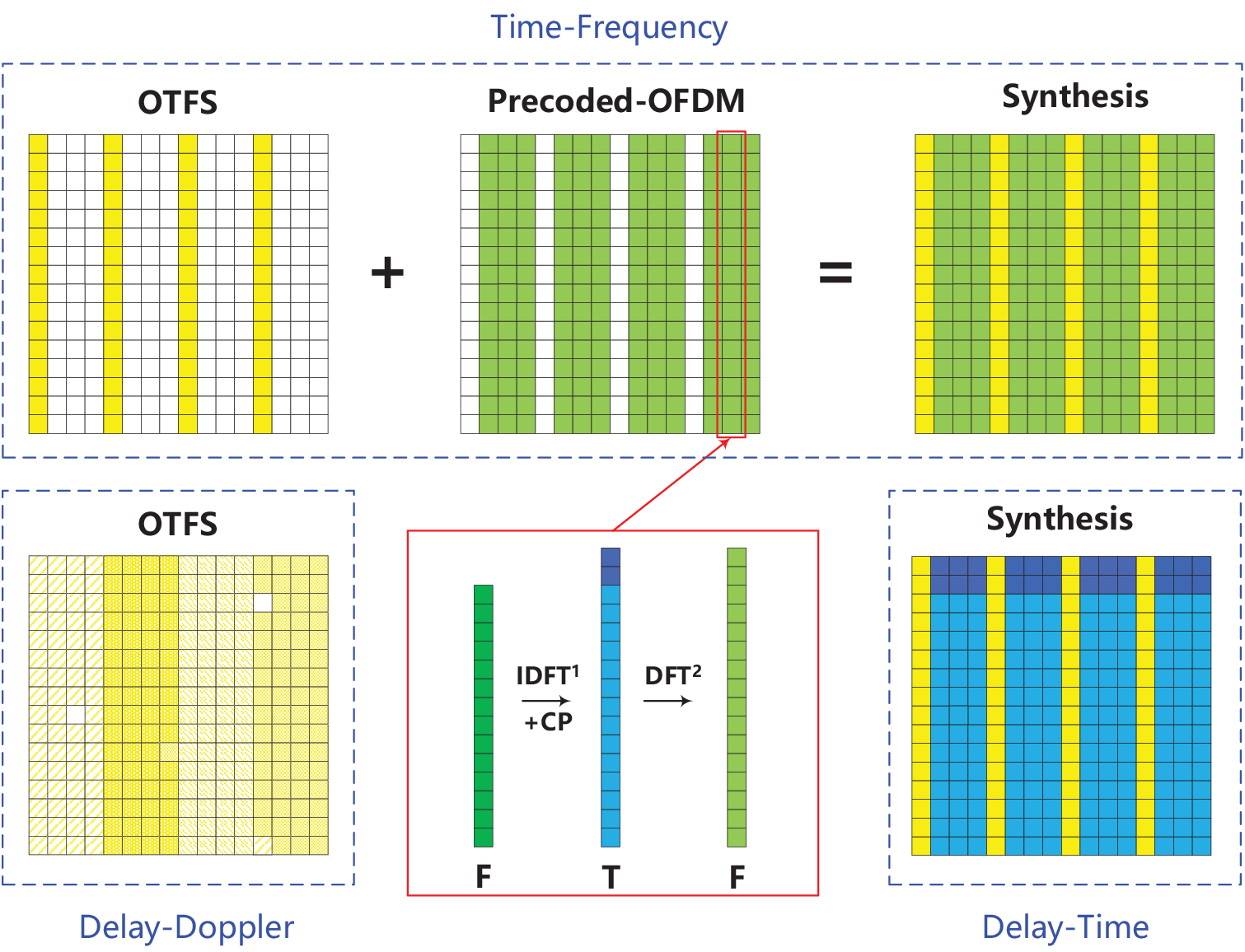}
  \caption[]{Illustration of time domain repetition.}
	\label{fig:fig0}
\end{figure}

Our implanting scheme is illustrated in figure \ref{fig:fig0}, in which the delay-Doppler domain modulation symbols repeat for four times along the Doppler dimension. After converted to time-frequency domain, three idle OFDM symbols out of four are leftover for OFDM transmission. We denote the three consecutive OFDM symbols as an OFDM slot. Depending on the service requirements, either one or multiple OFDM slots can be used for sending an OFDM data block. 

Assume that in time-frequency domain $N_{dd}$ OFDM symbols will be occupied by the OTFS and $N_{tf}$ will be occupied by the OFDM, we denote the OTFS data symbols as $\mathbf{S}_{dd}, \mathbf{S}_{dd}\in \mathbb{C}^{M \times N_{dd}}$ and the OFDM data symbols as $\mathbf{S}_{tf}, \mathbf{S}_{tf}\in \mathbb{C}^{M \times N_{tf}}$, where $N_{dd}+N_{tf}=N$. 

We apply a Doppler dimension extension to the OTFS data symbol matrix to obtain a $M\times N$ matrix, i.e., $\mathbf{1}_{1\times M}\otimes\mathbf{S}_{dd}$, where $\mathbf{1}_{1\times M}$ is a $1\times M$ vector with all $1$ entries. Then we apply the ISFFT to obtain the OTFS data symbols in time-frequency domain as follows.
\begin{equation}
\mathbf{X}_{dd} = \mathbf{F}_M(\mathbf{1}_{1\times M}\otimes \mathbf{S}_{dd})\mathbf{F}^H_N.
\label{eq:eq1}
\end{equation}

We apply a a time dimension zero interpolation to the OFDM data symbol matrix, i.e., start before the first column, we insert a zero columns every $N_{s}$ columns to obtain a $M\times N$ matrix for $N_{dd}$ times, where $N_{dd}\cdot N_{s}=N_{tf}$. Mathematically, it is expressed as the following equation.
\begin{equation}
\mathbf{X}_{tf} = \mathbf{S}_{tf}(\mathbf{I}_{N_{dd}}\otimes [\mathbf{0}_{N_{s}\times 1}, \mathbf{I}_{N_{s}}]),
\label{eq:eq2}
\end{equation}
where $\mathbf{0}_{N_{s}\times 1}$ is a $N_{s}\times 1$ vector with all $0$ entries, and $\mathbf{I}_{n}$ is the $n$-order unit matrix.

The transmit symbols in time-frequency domain is the sum of $\mathbf{X}_{dd}$ and $\mathbf{X}_{tf}$.
\begin{equation}
\mathbf{X} = \mathbf{F}_M(\mathbf{1}_{1\times M}\otimes\mathbf{S}_{dd})\mathbf{F}^H_N+\mathbf{S}_{tf}(\mathbf{I}_{N_{dd}}\otimes [\mathbf{0}_{N_{s}\times 1}, \mathbf{I}_{N_{s}}]),
\label{eq:eq3}
\end{equation}

The OTFS and OFDM symbols are orthogonally mapped in the time-frequency domain as (\ref{eq:eq3}). After obtaining the time-frequency domain resource mapping, we use the Heisenberg transform analogous to conventional OFDM to get the time domain signal.
\begin{align}
\mathbf{S} & = \mathbf{G}_{tx}\mathbf{F}^H_M\mathbf{X} \nonumber \\ 
& = \mathbf{G}_{tx}(\mathbf{1}_{1\times M}\otimes\mathbf{S}_{dd})\mathbf{F}^H_N \nonumber \\ 
& +\mathbf{G}_{tx}\mathbf{F}^H_M\mathbf{S}_{tf}(\mathbf{I}_{N_{dd}}\otimes [\mathbf{0}_{N_{s}\times 1}, \mathbf{I}_{N_{s}}]),
\label{eq:eq4}
\end{align}

We follow the sameline as in \cite{Raviteja2019embedded} where the rectangle pulse shaping is adopted, then the vector form of (\ref{eq:eq4}) can be written as,
\begin{equation}
\mathbf{s} = {\mathbf{F}^H_N\otimes\mathbf{I}^M}\mathbf{x}_{dd}+{\mathbf{I}^N\otimes\mathbf{F}^H_M}\mathbf{x}_{tf},
\label{eq:eq5}
\end{equation}
where 
$${\mathbf{x}_{dd} = \mathbf{vec}(\mathbf{F}_M(\mathbf{1}_{1\times M}\otimes \mathbf{S}_{dd})\mathbf{F}^H_N)},$$
$${\mathbf{x}_{tf}=\mathbf{vec}(\mathbf{S}_{tf}(\mathbf{I}_{N_{dd}}\otimes [\mathbf{0}_{N_{s}\times 1}, \mathbf{I}_{N_{s}}]))},$$
and $\mathbf{vec}(\cdot)$ denotes the column-by-column vectorization of a matrix. 

\subsection{OFDM Symbol Precoding}
Since the OFDM requires a CP to remove the ISI and convert the channel coupling effect from linear convolution to circular convolution to enable the single-tap FDE (frequency domain equalization),  while the OTFS does not. 
We propose the following precoding scheme to add the CP for the OFDM while keeping the same symbol time of the CP appended OFDM with the CP-free OTFS. 

We denote the original modulated symbols of the OFDM user as $\bar{\mathbf{s}}_{tf}, \bar{\mathbf{s}}\in \mathbb{C}^{M-M-L_{cp} \times N_{tf}}$. As illustrated in figure \ref{fig:fig0}, we first use a size $M-L_{cp}$ DFT to obtain the time domain signal, then add the CP and convert it back to frequency domain using a size $M$ IDFT. This process is equivalents to a precoding process as the follow,
\begin{equation}
\mathbf{s} = \mathbf{F}_M \mathbf{B}_{cp} \mathbf{F}^H_{M-L_{cp}} \bar{\mathbf{s}}_{tf},
\label{eq:eq6}
\end{equation}
where the precoding matrix is $\mathbf{F}^H_N \mathbf{B}_{cp} \mathbf{F}_{M-L_{cp}}$ and 
$\mathbf{B}_{cp} = \left[
\begin{array}{ccc}
\mathbf{0}_{L_{cp}\times (M-2L_{cp})} &  & \mathbf{I}_{L_{cp}}  \\
 & \mathbf{I}_{M-L{cp}} &   \\
\end{array}
\right]$ 
is the operator for appending CP \cite{9303350}.

After the precoding process, the OFDM symbol is protected by the CP in time domain while keeping the same symbol duration with the OTFS. In this way, the time domain channel consistency for the OTFS users is maintained and the transformation rule between time-frequency and delay-Doppler domain would not be violated. 
\begin{figure}
\centering
        \includegraphics[width=0.45\textwidth]{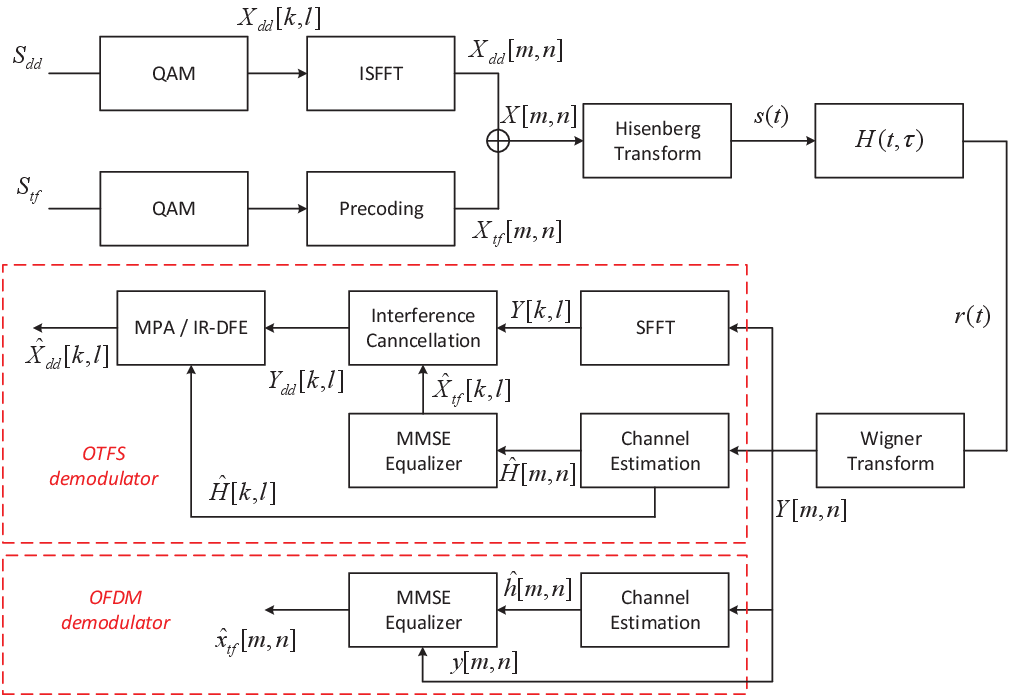}
        \caption{The structure of the transmitter and the receiver.}
    \label{fig:fig2}
\end{figure}

\section{Receiver Processing for Symbol Detection}
The proposed frame structure achieves a synthesis waveform where the OTFS and OFDM orthogonally multiplexed in the time domain, therefore it is expected to be easily separated and demodulated independently. However, the orthogonality will be disrupted after coupling with the multi-path delay channel and the ISI is introduced.

Starting form the received time-domain signal $r(t)$, the continuous time domain input-output relation can be written as,
\begin{equation}
r(t) = \int^{\tau_{max}}_{0}{h(\tau,t)s(t-\tau)d\tau},
\label{eq:eq7}
\end{equation}
where $h(\tau,t)$ is the continuous time-varying channel impulse response. The discrete time form of $r(t)$ after sampling is written as,
\begin{equation}
\mathbf{r}(q) = \sum_{l\in \mathcal{L}}{h(l,q)s(q-l)}.
\label{eq:eq8}
\end{equation}
where $\mathcal{L}$ represents the set of all path delays. 

The time domain input-output in (\ref{eq:eq8}) of the OTFS frame can be written in vector form as $\mathbf{r} = \mathbf{H}_t\mathbf{s}+\mathbf{w}$ where $\mathbf{H}_t \in \mathbb{C}^{MN\times MN}$ is the equivalent time domain channel matrix determined from path delay, Doppler, channel gain and shaping pulse in the same way as \cite{Raviteja2018interference}. We can substitute (\ref{eq:eq6}) in above equation and get the follows,
\begin{equation}
\mathbf{r} = \mathbf{H}_t({\mathbf{F}^H_N\otimes\mathbf{I}^M}\mathbf{x}_{dd}+{\mathbf{I}^N\otimes\mathbf{F}^H_M}\mathbf{x}_{tf})+\mathbf{w}.
\label{eq:eq9}
\end{equation}

Obviously, if the time domain equivalent channel matrix represented by $\mathbf{H}_t$ is not a diagonal matrix, then the mutual interference between the OFDM and OTFS is introduced and efficient interference mitigation schemes are demanded.

\subsection{OFDM Detection}
In the proposed structure, the OFDM symbol is appended with a CP to enable the single tap equalizer. A properly CP length which greater equal than the maximum delay spread will eliminate the ISI after detaching the CP. Hence the symbol-by-symbol detection is still applicable for OFDM user in the hybrid frame structure.

The received signal is firstly removed the CP and then converted to the 2D time-frequency grid using the Wigner transform, followed by a single tap FDE to decouple the channel. If the MMSE criterion is adopted, the OFDM detection can be written as,
\begin{equation}
\hat{\mathbf{x}}_{n} = (\mathbf{h}_{t}^{*}\circ (\mathbf{F}_{M-L_{cp}}\mathbf{B}^{-1}_{cp}\mathbf{r}_{n}))\oslash (\left|\mathbf{h}_{t}\right|^{2}+\sigma^2),
\label{eq:eq10}
\end{equation}
where $\sigma^2$ is the noise variance, $\mathbf{x}_{n}$ is the transmit data symbols on the $n$-th OFDM symbol and $\mathbf{r}_{n}$ is the corresponding received signal in time domain, $\mathbf{h}_{t}^{*} \in \mathbb{C}^{(M-L_{cp})\times 1}$ is the conjugate of the frequency domain channel.  
\subsection{OTFS Detection}
In contrast to OFDM, the OTFS user is prone to ISI in the absence of CP, which makes the symbol detection challenging. 
According to (\ref{eq:eq10}), a straightforward way to mitigate the ISI is directly equalizing $\mathbf{r}$ in the time domain upon obtaining the channel matrix $\mathbf{H}_t$. This kind of equalizers, such as $(\mathbf{H}_t^H\mathbf{H}_t+N_0\mathbf{I})^{-1}\mathbf{H}_t^H$ requires calculating the inverse of a $MN\times MN$ matrix, which is rather large and may be prohibited from practical applications.  

Recall that the entries of the delay-Doppler domain channel response matrix $\mathbf{H}_{dd}\in \mathbb{C}^{M\times N}$ is uniquely obtained defined by a set $<h_{l_i},\tau_{l_i},\nu_{l_i}>$, where we denote $h_{l_i}$, $\tau_{l_i}$ and $\nu_{l_i}$ as the as the channel gain, delay and Doppler respectively. Here $\tau_{l_i}$ and $\nu_{l_i}$ determines the position while $h_{l_i}$ determines the value of the channel response of the $l_i$-th propagation path.

As the channel response couples with the transmitted signal in a two dimensional convolution manner in the delay-Doppler domain, the corresponding time-frequency domain channel response couples with the transmitted signal in a element-wise product manner, if we assume ideal pulse shaping is adopted \cite{Raviteja2018lowcomplexity}. Although in practice we adopt the rectangular pulse and this relationship does not strictly holds, we can still apply a low complexity frequency-time domain equalizing to alleviate the ISI to enable a rough separation of the OFDM and the OTFS symbols. 
\begin{equation}
\hat{\mathbf{Y}} = (\mathbf{H}_{tf}^{*}\circ (\mathbf{F}_{M}\mathbf{R}))\oslash (\left|\mathbf{H}_{tf}\right|^{2}+\sigma^2),
\label{eq:eq11}
\end{equation}
where $\mathbf{r} = {\rm vec}(\mathbf{R})$ and $\mathbf{R} = {\rm vec}^{-1}(\mathbf{r})$.

Since most of the ISI is removed in the operation in (\ref{eq:eq11}), we can extract the OTFS component from the received signal by directly removing the OFDM data symbols in $\hat{\mathbf{Y}}$ and get the delay-Doppler domain symbols as follows,
\begin{equation}
\hat{\mathbf{Y}}_{dd} = \mathbf{F}^{H}_{M}
\left(\hat{\mathbf{Y}}
\left(\mathbf{I}_{N_{dd}}\otimes 
\left[
\begin{array}{cc}
1 &   \\
 & \mathbf{0}_{N_{s}}   \\
\end{array}
\right]
\right)
\right)
\mathbf{F}_{N},
\label{eq:eq12}
\end{equation}
where $\mathbf{0}_{N_{s}}$ is zero square matrix of dimension $N_{s}$.

After the above processing, conventional OTFS detection schemes such as the MPA or IR-DFE can be directly applied to $\hat{\mathbf{Y}}_{dd}$. 

\section{Interference Cancellation Protocol}
The time-frequency domain separation (TFDS) can be easily implemented but the approximation results in the residual ISI, which will degrade the symbol detection performance. In this section we provides an alternative approach to remove the ISI and discuss the corresponding protocols.
\subsection{The Interference Cancellation}
Suppose the OTFS user is allowed to demodulate the OFDM data under some protocols. The entries of the OFDM data symbol matrix $\hat{\mathbf{X}}_{tf}$ is estimated by the OTFS receiver analogous to (\ref{eq:eq9}) with a hard decision on exact modulated symbols. Based on the estimated $\hat{\mathbf{X}}_{tf}$ the interference cancellation can be done to the received signal. 

We can cancel the information in time domain directly as follows. Upon obtaining the equivalent time domain channel matrix $\mathbf{H}_{t}$ and estimated OFDM transmit data symbols matrix $\hat{\mathbf{X}}_{tf}$, we can write down the OFDM interfering term at the OTFS receiver as,
\begin{equation}
\Delta \mathbf{r} = \mathbf{H}_{t} {\rm vec}(\mathbf{F}^{H}_M\hat{\mathbf{X}}_{tf}),
\label{eq:eq13}
\end{equation}

Then we obtain the interference canceled OTFS delay-Doppler domain symbols as,
\begin{equation}
\hat{\mathbf{Y}}_{dd} = \mathbf{F}^{H}_{M}{\rm vec}^{-1}(\mathbf{r}-\Delta \mathbf{r})\mathbf{F}_{N}.
\label{eq:eq14}
\end{equation}

After the time domain interference cancellation (TDIC) or the TFDS mentioned in the previous section, conventional OTFS detection algorithms are followed to detect the modulated symbols. As in the proposed structure the OTFS frame covers a number of OFDM slots, therefore low complexity solutions are preferred. Enlightened by \cite{Raviteja2018lowcomplexity} and \cite{9293173}, the IR-DFE is adopted for OTFS symbol detection after interference cancellation, where the idea of maximum ratio combining is borrowed for SINR enhancement \cite{holter2002optimal}. 

\subsection{Protocols Design}
Suppose the OTFS and OFDM are adopted by different users. The symbol level interference cancellation at the receiver relies on the knowledge on the configurations of the OFDM users, such as modulation scheme and resource occupancy. 
\begin{figure}
\centering
        \includegraphics[width=0.45\textwidth]{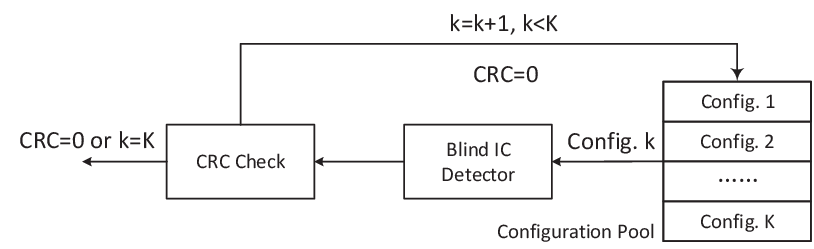}
        \caption{The blind interference cancellation procedure.}
    \label{fig:fig3}
\end{figure}
A straightforward way to indicate these configurations is to send a message from the transmitter to the receiver. Based on the collected information, the OTFS receiver could demodulate the OFDM symbols with the proposed interference cancellation scheme. However, if the OTFS and OFDM users are different individuals, information security problem may arise since the transmit symbols of the OFDM users are demodulated by someone else. Fortunately, this risk can be avoided since the actual transmitted bits could be encrypted and protected by the bit-level encoding.

Furthermore, in practical protocol there are only a few alternative MCS (modulation and coding schemes) and resource configurations, the OTFS user is enabled to perform blind interference cancellation to reduce the signaling overhead, which is illustrated in figure \ref{fig:fig3}. For example, the OFDM user can be forced to rate-match to one OFDM slot with limited choice of sizes, then the uncertainty on the resource occupancy is reduced. Meanwhile, a long term configuration pool of the MCS is provided with only a few candidates. Then with only a few hypothesis tests, the OTFS receiver could successfully guess the modulation order of the OFDM and cancel the interference. In such cases the signaling overhead is negligible.

\section{Numerical Results}
In this section, we provide the numerical evaluation. We consider a frame of size $N\times M$, where $N$ and $M$ correspond to the number of Doppler and delay taps in the delay-Doppler domain and the number of OFDM symbols and sub-carriers in the time-frequency domain. We assume that each of the OFDM symbols is allocated with the same power to maintain consistency in the power amplifier state, while ensuring a fair comparison between OTFS and OFDM. The channel delay model adopts the 3GPP extended vehicular A (EVA), extended pedestrian A (EPA), and extended typical urban (ETU) models, and the Doppler shift for the $i_{th}$ path follows a uniform distribution $U(0, \nu_{max})$ where $\nu_{max}$ is the maximum Doppler shift calculated based on the velocity and carrier frequency. The detailed system setup is shown in Table \ref{tab:tab0}.
\begin{table}
\begin{center}
\caption{System Parameters}
\label{tab:tab0}
\resizebox{\linewidth}{!}{
\begin{tabular}{|c||c|}
\hline
    Parameter & Value \\
\hline
    Carrier frequency & 28GHz\\ 
\hline
    Sub-carrier spacing & 60e3\\ 
\hline
    [$N$,$M$] & [16,512]\\
\hline
    [$N_{dd}$,$N_{tf}$] & [8,8]\\
\hline
    Delay and Power profile &  3GPP EVA, EPA, ETU\\
\hline
    Modulation & 16QAM\\ 
\hline
    Code & Low-density parity-check code (LDPC), rate = 0.5\\ 
\hline
    Velocity & 3, 30, 200, 300, 500km/h\\ 
\hline
\end{tabular}}
\end{center}
\end{table}

Both the performances of time-frequency domain separation and interference cancellation are evaluated. We investigate the bit error probability (BER) under perfect CSI and also the BER under imperfect CSI obtained by delay-Doppler domain channel estimation. In both cases, the raw BER (i.e., the BER without channel coding) and the coded BER are provided.

\begin{figure}
\centering
\includegraphics[width=0.45\textwidth]{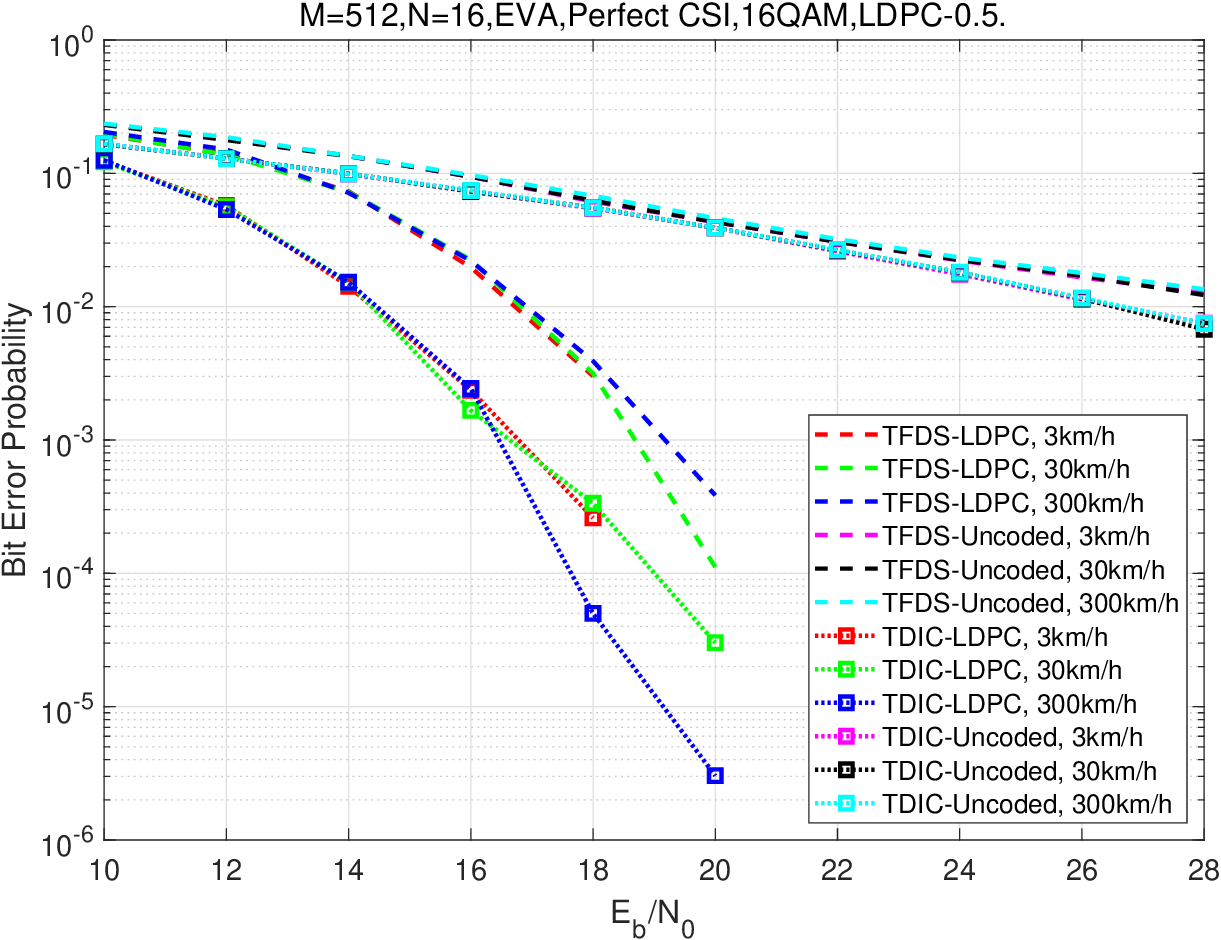}
  \caption[]{\label{fig:Num0}{} TDIC outperforms TFDS with perfect CSI.}
\end{figure}

\begin{figure}
\centering
\includegraphics[width=0.45\textwidth]{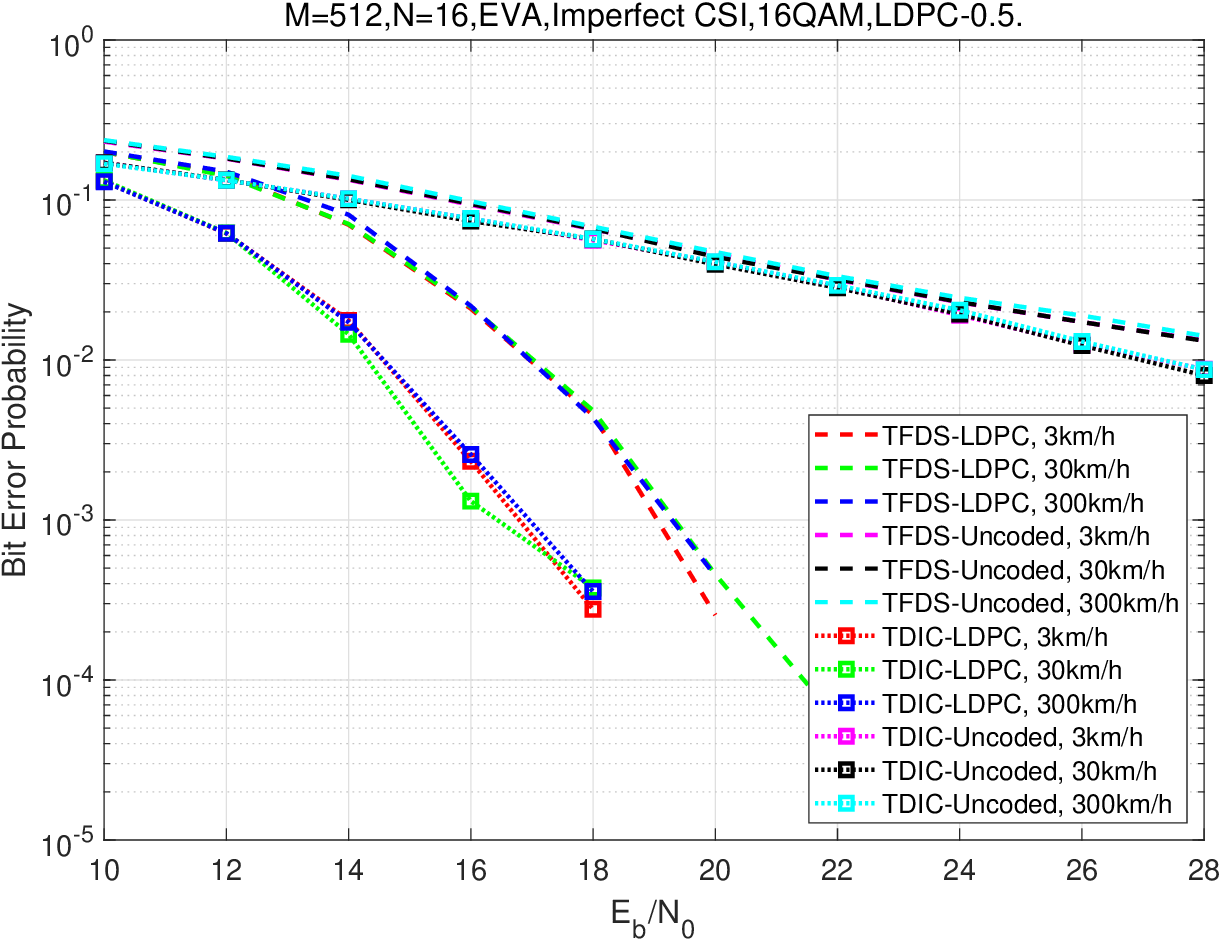}
 \caption[]{\label{fig:Num1}{} TDIC outperforms TFDS with imperfect CSI.}
\end{figure}

We first demonstrate the role of interference cancellation via a comparison of TFDS and TDIC in Figures \ref{fig:Num0} and \ref{fig:Num1}. It is observed in both coded and uncoded systems that TDIC outperforms TFDS in all cases. However, TFDS is still an option to consider for its low complexity and overhead.

\begin{figure}
\centering
\includegraphics[width=0.45\textwidth]{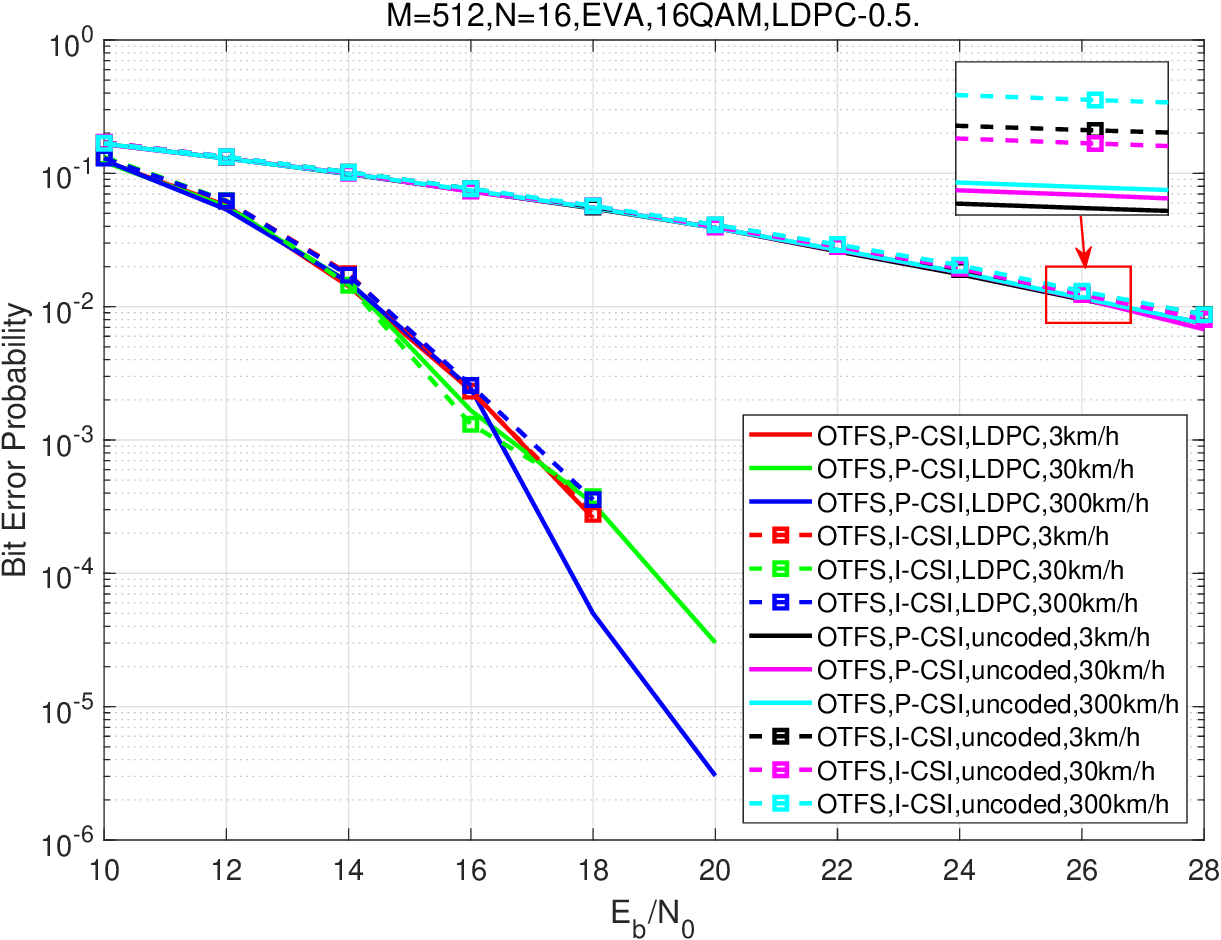}
  \caption[]{\label{fig:Num2}{} TDIC performance evaluation.}
\end{figure}

We then demonstrate the performance of the TDIC in both perfect and imperfect CSI in figure \ref{fig:Num2}. The performance of TDIC with perfect CSI is slightly better than that with imperfect CSI in terms of raw BER. However, this marginal advantage vanishes after introducing the channel coding. Therefore we can conclude that the proposed TDIC is robust in practical communication systems.

\begin{figure}
    \centering
        \includegraphics[width=0.45\textwidth]{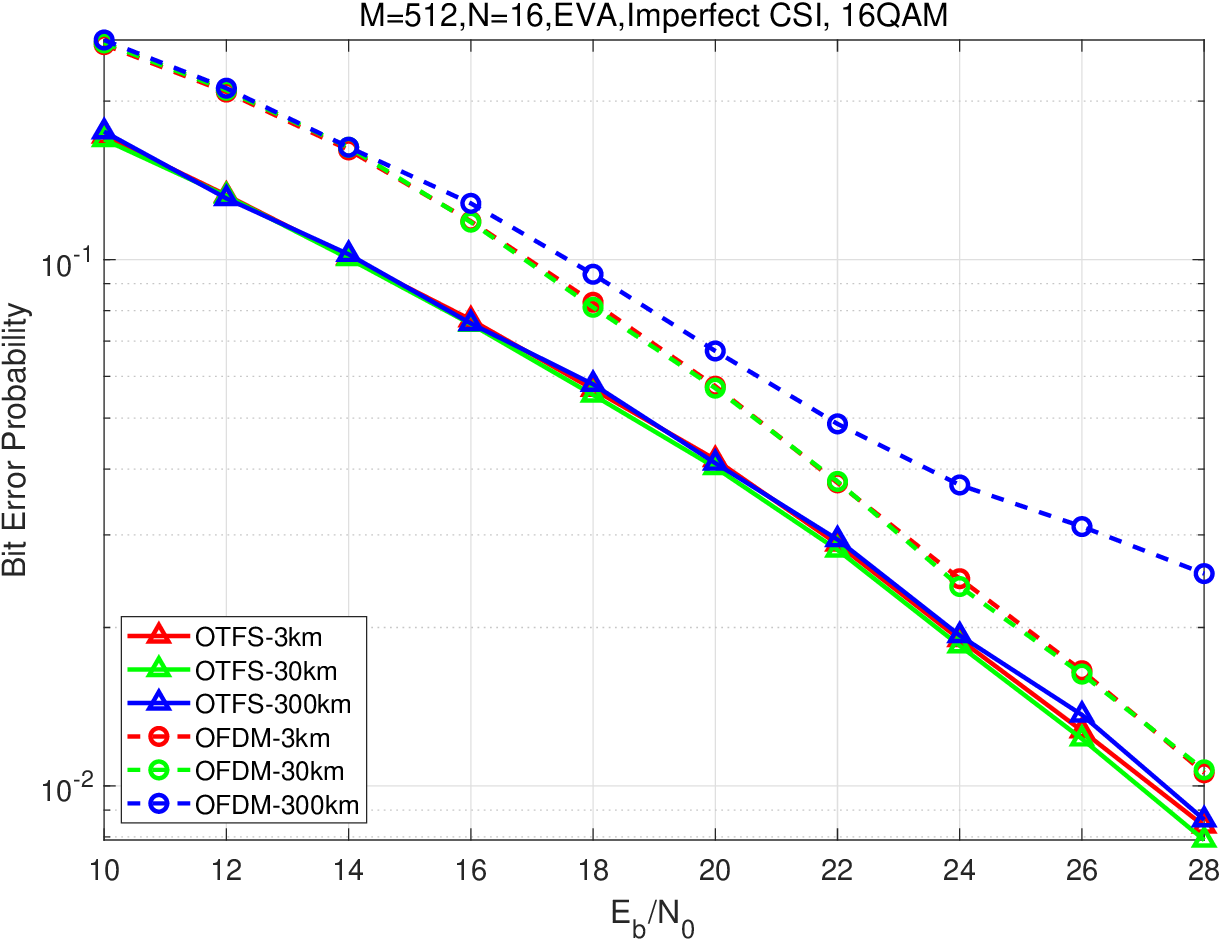}
        \caption{The OTFS outperforms the OFDM without channel coding.} 
    \label{fig:Num4}
\end{figure}

\begin{figure}
    \centering
        \includegraphics[width=0.45\textwidth]{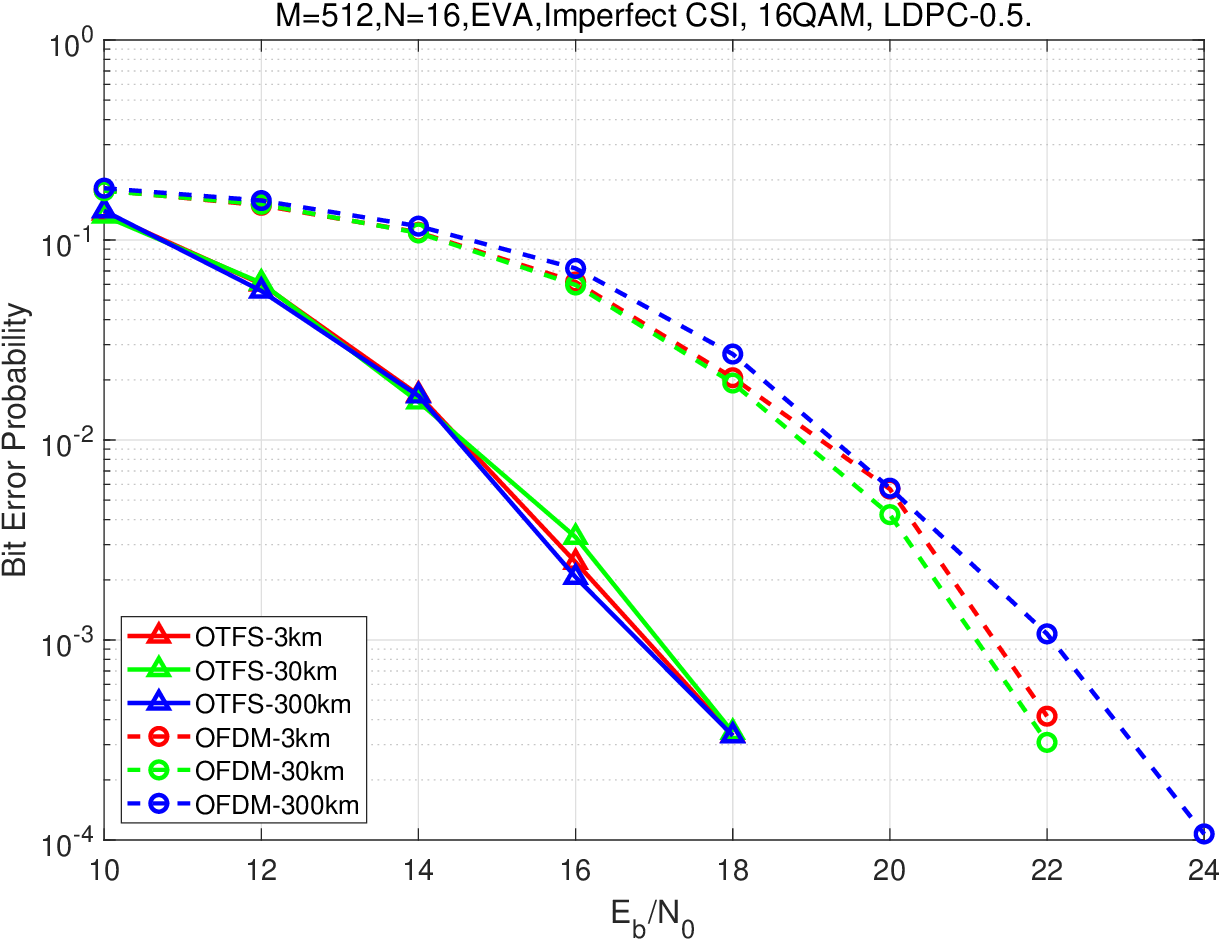}
        \caption{The OTFS outperforms the OFDM with channel coding.} 
    \label{fig:Num5}
\end{figure}

Figures \ref{fig:Num4} and \ref{fig:Num5} illustrate the performances of both OTFS and OFDM in the hybrid frame structure with perfect and imperfect CSI, respectively. At a speed of $3$km/h, the raw BER of OTFS is recorded as $1\%$ at $27$dB, while OFDM achieves the same BER at $28$dB. As the speed increases to $30$km/h, the raw BER of both OTFS and OFDM remains roughly the same as that at $3$km/h. At a speed of $300$km/h, the raw BER of OTFS is recorded as $0.02$ at $22$dB, while OFDM is $4$dB worse. This observation is consistent with the common knowledge that OFDM is prone to high Doppler, while OTFS is not.

Aided by LDPC coding, the performances of both OFDM and OTFS are significantly improved, as shown in Figure \ref{fig:Num5}. If the OFDM user occupies multiple OFDM symbols, cross-symbol diversity could be introduced by the channel coding. Therefore, we observe that OFDM enjoys more coding gain in high Doppler scenarios. The performance gap between coded OTFS and OFDM in the hybrid frame structure is around $4$ to $6$ dB using the setup in Table \ref{tab:tab0}.

\begin{figure}
    \centering
        \includegraphics[width=0.45\textwidth]{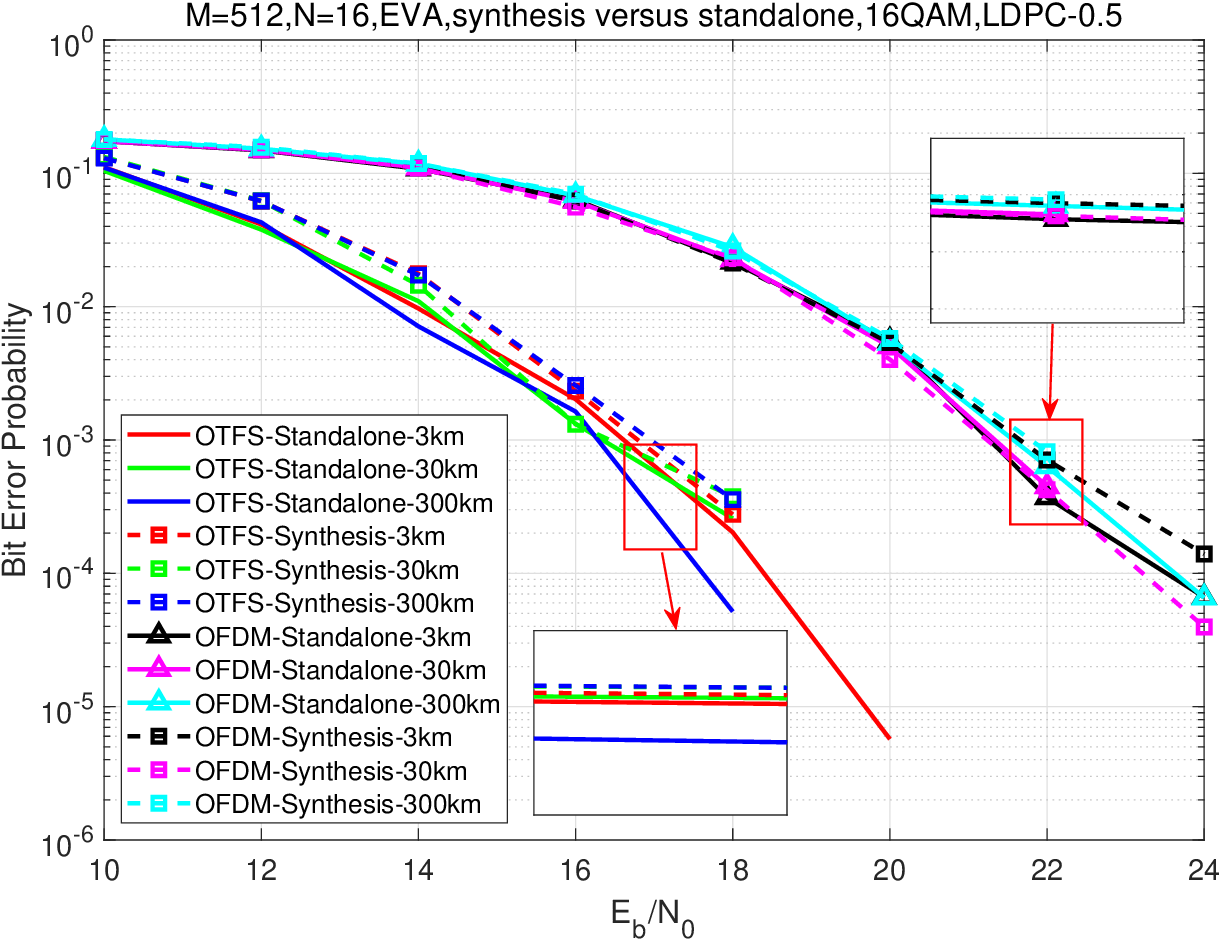}
        \caption{The performance of the hybrid frame structure and standalone ones with channel coding.} 
    \label{fig:Num6}
\end{figure}

Figure \ref{fig:Num6} illustrates the performance gap between the proposed hybrid frame structure and the standalone ones. It is observed that the performance of the OFDM in the hybrid frame structure is comparable to the standalone one due to the adequate protection by the CP. Furthermore, the TDIC-based OTFS successfully removes the ISI, with only a slight performance degradation.

\begin{figure}
    \centering
        \includegraphics[width=0.45\textwidth]{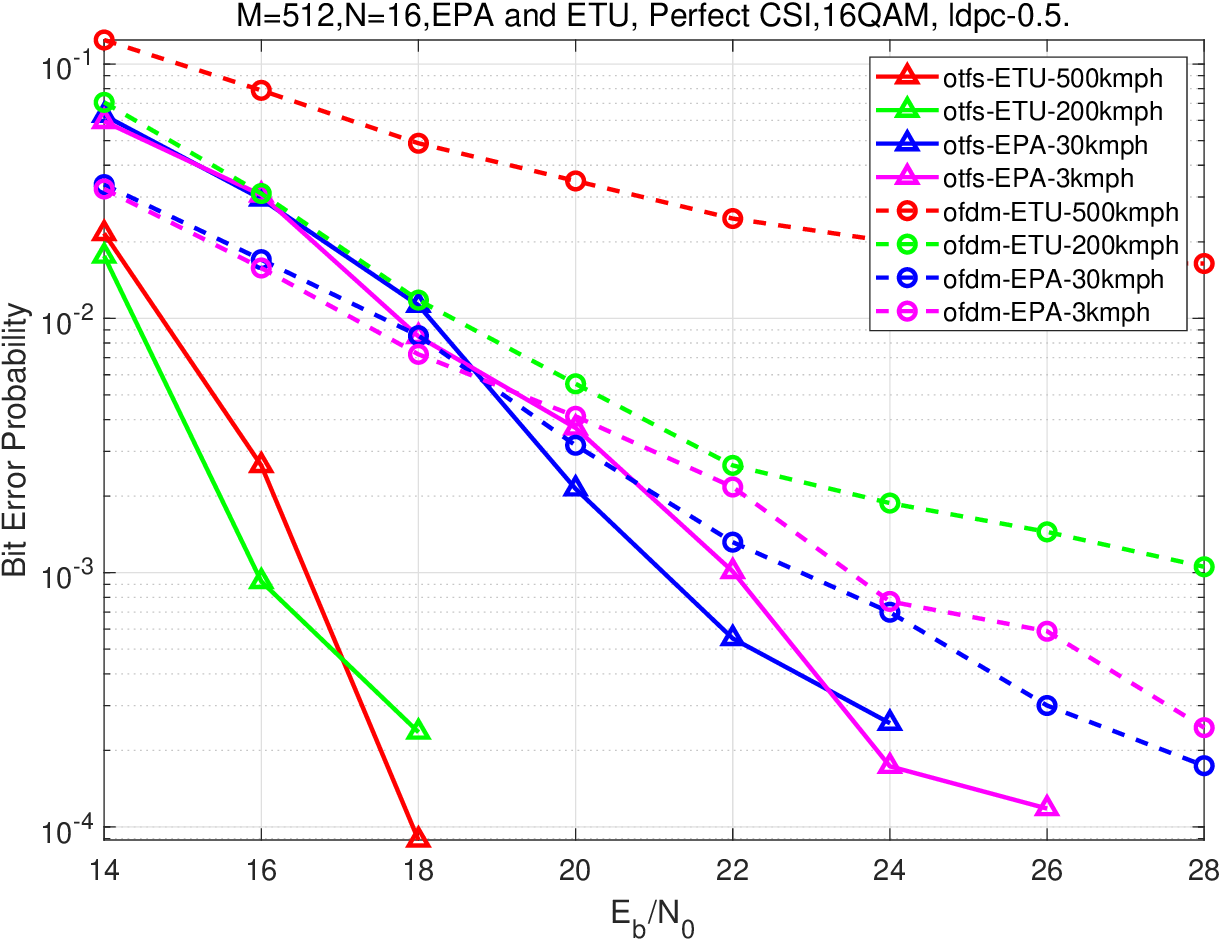}
        \caption{The performance of the hybrid frame structure in EPA and ETU models.} 
    \label{fig:Num7}
\end{figure}

Figure \ref{fig:Num7} demonstrates more results in EPA and ETU channel models, followed by the same observation as in Figure \ref{fig:Num6}. Therefore we can conclude that the proposed frame structure has a wide range of channel adaptability.

\section{Conclusion}
In conclusion, a hybrid frame structure consisting of OTFS and OFDM is studied in this paper. Based on the proposed transceiver design, users with different waveforms achieve co-existence where the mutual interference can be mitigated by the proposed interference cancellation scheme. The numerical results show the ISI is effectively mitigated by the TDIC where the performance difference between the the standalone OTFS or OFDM and the ones within the hybrid frame is less than $0.5$dB at the same BER. 
The proposed design provides a practical solution to accommodate multi-task-oriented communication in 6G, where the users with different numerology work simultaneously for different communication tasks to better exploit their advantages, and this framework can also be extended to other OTFS and OFDM variants for particular scenario, with a suitable symbols detector design. 

\bibliographystyle{IEEEtran}
\bibliography{fmtc_TWC}
\end{document}